\documentclass[aps,prd,preprint,showpacs]{revtex4}
\usepackage{amssymb,amsmath,graphicx}
\usepackage{verbatim,color,ulem}

\newcommand{\be}{\begin{equation}}
\newcommand{\ee}{\end{equation}}
\newcommand{\ba}{\begin{eqnarray}}
\newcommand{\ea}{\end{eqnarray}}

\newcommand{\dcom}[1]{}
\newcommand{\dnote}[1]{}

\newcommand{\gsim}{\raise.3ex\hbox{$>$\kern-.75em\lower1ex\hbox{$\sim$}}}
\newcommand{\lsim}{\raise.3ex\hbox{$<$\kern-.75em\lower1ex\hbox{$\sim$}}}

\begin{document}
\title{Geodesic Motion of Neutral Particles around a Kerr-Newman Black Hole}
\author{Chen-Yu Liu}
\email{ef850502@gmail.com}
\author{Da-Shin Lee}
\email{dslee@mail.ndhu.edu.tw}
\author{Chi-Yong Lin}
\email{lcyong@mail.ndhu.edu.tw}
\affiliation{Department of Physics, National Dong Hwa University,
Hualien, Taiwan, R.O.C.}

\date{\today}

\begin{abstract}
We examine the dynamics of a neutral particle around a Kerr-Newman black hole, and in particular focus on the effects of the charge of the spinning black hole on the motion of the particle.  We first consider the innermost stable circular orbits (ISCO) on the equatorial plane. It is found that the presence of the charge of the black hole leads to the  effective potential of the particle with stronger repulsive effects as compared with the Kerr black hole. As a result, the radius of ISCO decreases as charge $Q$ of the black hole  increases for a fixed value of  black hole's angular momentum $a$. We then consider a kick on the particle from its initial orbit out of the equatorial motion. The perturbed motion of the particle will eventually be bounded, or unbounded so that it escapes to spatial infinity. Even more, the particle will likely be captured by the black hole. Thus we analytically and numerically determine the parameter regions of the corresponding motions, in terms of the initial radius of the orbital motion  and the strength of the kick. The comparison will be made with the motion of a neutral particle in the Kerr black hole.
\end{abstract}


\maketitle

\section{Introduction} \label{sec1}
The phenomena of accretion disks and jets formation will  be the hallmark in observing the black hole~\cite{wheeler,hartle}. The energetic jets are created by the matter in accretion disks, and {are} powered perhaps from dynamic interactions within accretion disks or from  processes  associated with black holes. Relativistic jet formation may also explain observed gamma-ray bursts. Nevertheless, the {matter particles} in the surrounding of the black hole are significantly influenced by the gravitational pull. The investigation  of the motion of particles near the black hole certainly is of great importance  not only to  explore the spacetime structure of the black hole, but also to find the appropriate {mechanism} for the formation of accretion disks and jets.
In addition to the geodesic motion of test particles in ordinary general relativity, the interest of the subject has also gained attention in the field of higher dimensional gravity theories, more specifically  in connection to braneworld models ~\cite{Aliev:2005bi,Stuchlik:2008fy,Blaschke:2016uyo}.

The dynamics of particles orbiting around the black hole has been extensively studied {for}  Schwarzschild and Kerr black holes~\cite{wheeler,hartle,Pugliese:2011xn}. The exploration of the circular motion of the particle is also extended to  Reissner-Nordstr\"om  black holes~\cite{RUF1,RUF2}.  Recently, observational evidence for a correlation between jet power and black hole spin has been discovered~\cite{Nara}.  This finding is consistent with numerical simulations, showing that powerful jets
are generated by extracting energy from a spinning black hole together with the surrounding  magnetic field~\cite{Tch}.
These discoveries certainly {motivated more investigations on}  the evolution of particles in
weakly magnetized black holes. {For example, in the presence of weak magnetic field, the motion} of a charged particle near a Schwarzschild black hole is analyzed in~\cite{FRO1,FRO2,ZAH}, {while
around a Kerr black hole the charged particle can undergo chaotic motion}~\cite{ZAH1,RYOS,NAK,TAK,HUS}.
 Nevertheless, the Kerr-Newman metric represents a generalization of the Kerr metric, and describes spacetime in the exterior of a rotating charged black hole where, apart from
gravitation fields, both electric and  magnetic fields exist intrinsically from the black hole.

Although one might not expect that astrophysical black holes have a {large residue} electric charge,
{some} accretion scenarios were {proposed to investigate} the possibility of the spinning charged back holes~\cite{Damo}.
Thus, it is still of great interest to extend the {previous} studies to a Kerr-Newman black hole.
%
The geodesic motion of a test particle in the Kerr-Newman spacetime have been investigated. 
In particular, in~\cite{Pugliese:2013zma}, a detailed analysis of the orbital circular motion of electrically neutral test particles on the equatorial plane of the Kerr-Newman spacetime in the cases of black hole and naked singularity sources was presented.
The types of geodesic motion also include spherical and nonspherical orbits of test charges~\cite{RUF4}, and possible capture of particles from plunge orbits by a black hole~\cite{Young} (Also see~\cite{Sharp} and references therein). The general features of the radial motion, and the motion of charge particles along the axis of symmetry and in the equatorial plane, or  the special case of free-infalling particles were discussed in~\cite{Bicak1,Bicak2,Bicak3}.
One way to explore  the interplay between gravitational and
electromagnetic fields of a Kerr-Newman black hole is to consider
off-equatorial orbits of a charged particle  near the black hole~\cite{Kovar}. It is found that the stable off-equatorial orbits do not exist  above the outer horizon of the black
hole.
The analytical solutions of the
geodesic equations in the black hole spacetime were also found respectively in the Schwarzschild~\cite{Hagi}, Reissner-Nordstr\"om~\cite{Gru}, and Kerr-Newman~\cite{Hack} black holes. In particular, the paper~\cite{Hack} provided a comprehensive  classification of the orbits in radial and colatitudinal directions  for a large variety of orbit configurations of the charge particle in a Kerr-Newman black hole. Additionally, the analytical solutions of geodesic motion
for all types of orbits were presented in terms of elliptic functions dependent
on the so-called Mino time~\cite{Mino}.
According to these studies we can examine how the
 effects from electromagnetic fields intrinsically from the black hole  affect the dynamics of the surrounding particles.

 In this work, we will focus on the dynamics of a neutral  particle orbiting around a Kerr-Newman black hole, mainly following
 the strategy developed by~\cite{ZAH1}.  Most of results in~\cite{ZAH1} for a Kerr black hole will be generalized to a spinning charged black hole. { The effects of  electromagnetic fields associated with the Kerr-Newman black hole on a neutral particle are solely through the gravitational influence.
Here we {first consider} the so-called innermost stable circular orbits (ISCO) of the particle on the equatorial plane ($\theta=\pi/2$).  We  explore, in particular,  the effects of the black hole charge on the radius of ISCO via the appropriate effective potentials~\cite{Pugliese:2013zma}.
 Afterwards, the particle, initially moving around some particular orbit, is {perturbed by} a kick along the $\theta$ direction.
 %
 In this situation of possible astrophysical relevance, the kick could be given by another particles and/or photons.
 Then, the initial radius of the orbit and an amount of the kick form a two-dimensional parameter space.  We provide an analytical  analysis {in the same fashion as}~\cite{ZAH} to identify the {regions in the parameter} space, {that corresponds} to the different asymptotic behaviors of the particle.

 Although these trajectories can be described by means of the analytical solutions in the Mino time~\cite{Hack}, here we would like to explore them in the proper time instead. The nontrivial transformation between the Mino time and the proper time may suggest us to study the associated equations of motion directly in the proper time.
 Also in~\cite{Hack}, in spite of the fact that all possible trajectories were discussed, however in this work, we mainly focus on the above-mentioned situation  in more details.  The effects of the charge of a spinning black hole on the final states of the kicked particle can be understood analytically later by studying the appropriate effective potential~\cite{ZAH}.
  Finally, these analytical results will be reexamined by solving numerically the dynamical equations of the particle in
 the Kerr-Newman background with the corresponding initial conditions directly in the proper time. Same trajectories are believed to obtain as compared with the analytical solutions after the transformation from the Mino time to the proper time is carried out.


\section{Circular motion of a neutral particle around Kerr-Newman black hole} \label{sec2}

The {spacetime outside a} black hole with the gravitational mass $M$, charge $Q$, and  angular momentum per unit mass $a=J/M$ is described
 by Kerr-Newman metric:
\ba
 {ds}^2 &=& g_{\mu\nu} dx^\mu dx^\nu  \nonumber \\
&= & -\frac{ \left(\Delta -a^2 \sin^2\theta \right)}{\Sigma } {dt}^2   + \frac{ a   \sin ^2\theta \left(Q^2-2 M
   r\right)}{\Sigma } ({dt}{d\phi+ d\phi dt)}  \nonumber\\
&&\quad\quad+  \frac{\Sigma}{\Delta} dr^2 +\Sigma{\, d\theta}^2 +\frac{ \sin ^2 \theta}{\Sigma} \left(( r^2+a^2)^2 -a^2 \Delta  \sin
   ^2\theta \right) {d\phi}^2 \, , \label{KN_metric}
   \ea
where
\be
\Sigma=r^2+a^2\cos^2\theta \, , \quad  \Delta=r^2+a^2+Q^2-2 M r\,.
\ee
The  event horizon $R_{H}$ {can be found by solving $\Delta(r)=0$, and is given by}
\be
R_{H}=M+\sqrt{M^2-(Q^2+a^2)}\,
\ee
with $M^2 > Q^2+ a^2$.
Now we consider a test neutral particle with mass $m$ moves in  {such a spacetime.}  Its {Lagrangian}  is simply  {given by}
\be
L=-m \, \sqrt{- g_{\mu \nu} \, u^{\mu} u^{\nu}}  \, , \label{L_Nparticle}
\ee
where $u^{\mu}\equiv {dx^{\mu}}/{d\tau}$ is the four-velocity, {and $\tau$ is} proper time.
Because the metric of Kerr-Newman black hoke  has no explicit $t$ and $\phi$ dependence,  they  are cyclic coordinates in the
{Lagrangian}.  The respective Killing vectors, $ \xi^{\mu}_{(t)}$ and $ \xi^{\mu}_{(\phi)}$,   are given by
\be
\xi^{\mu}_{(t)} = \delta^{\mu}_t \, , \quad \quad\quad  \xi^{\mu}_{(\phi)}=\delta^{\mu}_\phi \, .
\ee
 Then, the associated  conserved quantities, namely energy  and azimuthal angular momentum per unit mass,  along a geodesic,  can be constructed by the above Killing vectors as well as the four-momentum $p_\mu =m u_\mu$ of the particle:
\be
 \varepsilon = - p_{\mu} \, \xi^{\mu}_{(t)}/m \, , \quad\quad
    \ell =  p_{\mu}\, \xi^{\mu}_{(\phi)}/m \, .
   \ee
 {Additionally, there exists a third constant of motion} (the Carter constant), defined by~\cite{ZAH}
 \be
 \kappa = u_\mu u_\nu K^{\mu \nu} -(\ell- a \varepsilon)^2 \, ,
 \ee
 where
 \begin{eqnarray}
 K^{\mu \nu}&=& \Delta [k^{ \mu } q^{\nu }- k^{ \nu } q^{\mu }] + r^2 g^{\mu \nu} \, , \nonumber\\
 q^{\mu} &=& \frac{1}{\Delta} [(r^2+a^2) \delta^{\mu}_t+\Delta \delta^{\mu}_r + a \delta^{\mu}_\phi ] \, ,\nonumber\\
 k^{\mu} &=& \frac{1}{\Delta} [(r^2+a^2) \delta^{\mu}_t-\Delta \delta^{\mu}_r + a \delta^{\mu}_\phi ] \, .
 \end{eqnarray}
 Using these three constants of motion and the normalization condition $u_{\mu} u^{\mu}=-1${, we are able to write down the equations of motion for the} components of $u^{\mu}$  in terms of $\varepsilon$, $\ell$, $\kappa$:
\begin{eqnarray}
\dot{t} &=& \varepsilon - \frac{(Q^2 - 2 M r)\, ((a^2+r^2) \varepsilon - a \ell)}{ \Delta \, \Sigma} \, , \\
\dot{\phi} &=& \frac{\ell}{ \Sigma \sin^2 \theta}-\frac{ a \,( (Q^2- 2 M r) \varepsilon + a \ell)} { \Delta \, \Sigma} \, ,\\
\Sigma^2 \dot r^2 &=& [ (r^2+a^2)\varepsilon- a \ell]^2-\Delta [r^2+\kappa+(\ell- a \varepsilon)^2] \, ,\label{rdot}\\
\Sigma^2 \dot \theta^2 &=& \kappa + (\ell- a \varepsilon)^2 - a^2 \cos^2 \theta -\bigg( a\varepsilon \sin\theta-\frac{\ell}{\sin\theta}\bigg)^2 \, . \label{thetadot}
\end{eqnarray}
The over  dot means the derivative with respect to the proper time $\tau$.

We restrict {the} particle {to move on the equatorial plane of} the  black hole
 by choosing $\theta={\pi}/{2}$, and $\dot{\theta}=0$ so that $\kappa=0$ from~(\ref{thetadot}).  The equation of the motion along the radial direction in~(\ref{rdot}) thus becomes
\be
\Sigma^2 \dot{r}^2 = r^4 \varepsilon^2 + 2 a \ell \varepsilon (Q^2 - 2 M r)+ a^2(\varepsilon^2 (r^2+2Mr-Q^2)- r^2) - (\Delta -a^2)(r^2 + \ell^2) \, . \label{rdotsquare}
\ee
The equation~(\ref{rdotsquare}) allows us to  define the effective potential $V_{\rm eff}$ by {requiring}
 \be
 \frac{1}{2}  \dot{r}^2 + V_{\rm eff}(r, {\bf \alpha} ) =\frac{\varepsilon^2-1}{2} \, ,
 \ee
where ${\bf \alpha}$  denotes a collection of the parameters of the Kerr-Newman black hole and the test neutral particle, namely $\alpha =( M,Q,a,\varepsilon, \ell)$
\be \label{v_eff}
V_{\rm eff} (r,\alpha) = -\frac{ M }{r}+\frac{\ell^2+ (1- \varepsilon^2) a^2 + Q^2}{2 r^2}  -\frac{ M  (\ell- a \varepsilon)^2 }{ r^3} +\frac{Q^2 (\ell -a \varepsilon)^2}{r^4} \, .
\ee
In the limits of $Q=0$ and $Q=0,\;a=0$, the effective potential reduces to the {ones for the} Kerr and Schwarzschild black holes respectively~\cite{wheeler,hartle}.
There are several distinct {classes of} motion, depending on whether the black hole spin and azimuthal angular momentum {of the particle} are aligned $\ell a>0$ or oppositely aligned $\ell a<0$.  Without losing generality, we keep $\ell$ positive and choose $a$ to be either positive or negative.   In the Kerr-Newman black hole,  the effective potential has the $Q^2$ dependence, and thus the sign of $Q$ will not be relevant. {Thus we choose}  $Q>0$ in {the present} work.  The non-zero charge of the black hole seems to give repulsive effects  to the particle as seen from its contributions to the centrifugal potential of the $1/r^2$ term as well as the relativistic  correction of the $1/r^4$ term. These repulsive effects will shift the radius of the stable circular motion toward the black hole, as will be shown later.
Also, the value of the $\varepsilon$ certainly determines whether or not the particle has chance to escape from the black hole to  spatial infinity.

To determine the circular motion, one may define $R (r)$ by the right hand side of the equation~(\ref{rdotsquare}),
\be
R(r) = r^4 \varepsilon^2 + 2 a \ell \varepsilon (Q^2 - 2 M r)+ a^2(\varepsilon^2 (r^2+2Mr-Q^2)- r^2) - (r^2+Q^2-2 M r)(r^2 + \ell^2) \,
  \, . \label{R}
\ee
{Thus its zero tells us the radius, say $r_{o}$, at which the velocity along the $r$ direction vanishes}.  The existence of the  circular orbits requires that the first derivative {of $R$} with respect to $r$ {at $r=r_{o}$} vanishes. {This means zero-acceleration} along the $r$ direction. {Moreover the condition}  $R''(r_o) \geq 0$ {guarantees that circular motion is stable}.
 The {first two} conditions give:
\be
r^4 \varepsilon^2 + 2 a \ell \varepsilon (Q^2 - 2 M r)+ a^2(\varepsilon^2 (r^2+2Mr-Q^2)- r^2) - (r^2+Q^2-2 M r)(r^2 + \ell^2) = 0 \, ,
\ee
\be
r[(3 M r -Q^2)+ 2 r^2 (\varepsilon^2 - 1)]+\ell((M-r)\ell-2 a M \varepsilon)+a^2(M \varepsilon^2 + r(\varepsilon^2 -1)) = 0 \, ,
\ee
from which, $\varepsilon_o$ and $\ell_o$ {for} this circular orbit {with} a given radius $r_o$ are determined straightforwardly as
\ba
\varepsilon_o = \frac{ a \sqrt{ M r_o-Q^2} + (Q^2+r_o^2-2 M r_o)}{ r_o \sqrt{ 2 Q^2 + r_o^2- 3 M r_o + 2 a ( M r_o-Q^2)^{1/2}}} \, ,\label{e_o} \\
\ell_o=\frac{ a (Q^2 -2 M r_o)+ ( a^2 + r_o^2) \sqrt{M r_o -Q^2}}{ r_o \sqrt{ 2 Q^2 + r_o^2- 3 M r_o + 2 a ( M r_o-Q^2)^{1/2}}} \, .\label{l_o}
\ea
{That is, we must choose $\varepsilon_{o}$ and $\ell_{o}$ given the equations above in order for a neutral particle to take a circular orbit of radius $r_{o}$ around the Kerr-Newman black hole}.
Apparently, $r_o$ can not be arbitrarily small.
When the radius $r_o$ is smaller than the so-called the radius of the inner most stable circular orbit, $r_{\rm ISCO}$ given by $R''(r_o)=0$, the circular motion becomes unstable ~\cite{Blaschke:2016uyo,Pugliese:2013zma}.
{Thus $r_{\rm ISCO}$ satisfies}
\be
(6~r_{\rm ISCO}^2+a^2)(\varepsilon_o^2-1)+6 M r_{\rm ISCO}-\ell_o^2-Q^2 = 0 \, . \label{isco}
\ee
%
Substituting the expressions ~(\ref{e_o}) and ~(\ref{l_o}) into  ~(\ref{isco}) we arrive at the following polynomial equation for the radius of the circular orbits, also obtained in~\cite{Aliev:2005bi},
\be
M r_{\rm ISCO}(6Mr_{\rm ISCO}-r^2_{\rm ISCO}-9Q^2+3a^2)+
4Q^2(Q^2-a^2)-8a(M r_{\rm ISCO}-Q^2)^{3/2}\;=\;0\,.
\ee
In Fig.~\ref{iscovsavq} we plot the behavior of $r_{\rm ISCO}$ {for} various combinations of parameters $a$ and $Q$.

\begin{figure}
  \centering
  \includegraphics[scale=0.45]{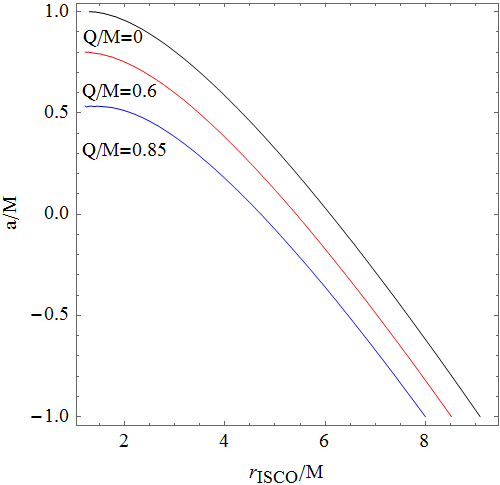}
  \caption{The dependence of the radius $r_{\rm ISCO}$ on black hole's angular momentum per unit mass ${a}$ and charge per unit mass ${Q}$ . } \label{iscovsavq}
\end{figure}

For a given $Q$, {in the case of $a>0$ the radius of the ISCO} decreases with the increase of $a$  whereas the radius {for $a<0$} increases  when
the absolute value of $a$ increases. In particular, {we highlight the following three cases for the neutral black hole,} $Q=0$. {(A) if the black hole does not rotate, i.e., Schwarzschild black hole ($a=0$), we have} $r_{\rm ISCO}=6 M$, {while (B) in extreme limit $a=1$, we find $r_{\rm ISCO}=1 M$ when the particle rotates in the same orientation as the Kerr black hole, but (C) $r_{\rm ISCO}=9 M$ in the opposite orientation}.
For a nonzero $Q$, the ISCO radius generally is smaller than in the $Q=0$ case for a given $a$.  This {results from the fact that} the charge of the black hole seems to give  more repulsive effects on the particle as compared with the $Q=0$ case, as seen in the effective potential~(\ref{v_eff}).
In {comparison with (A)}, when $Q=M, a=0$, the ISCO radius is $r_{\rm ISCO}= 1 M$.
All circular motion corresponds to  its energy $\varepsilon <1$~\cite{Pugliese:2013zma}. In the next section, we {will} explore the {case that the particle, which initially orbits on the equatorial plane, experiences an impulse force in the direction normal to the plane}.

\section{Final states of the particle kicked off from an initial circular motion} \label{sec3}
{In general, the coupled equations of motion for a neutral particle moving around a Kerr-Newman black hole are to complicated to solve in the proper time. In order to} make analytic investigation possible, we consider {a special case that} the particle {is knocked out of its original circular motion on the} equatorial  plane by an impulse force along the $\theta$ direction. {Thus the circular motion of the particle and the influence of the kick can be effectively summarized by the} four-velocity {at the moment when the force acts on the particle,}
\be
u_{i \, \nu} = (-\varepsilon, 0, r_o^2 \dot\theta_k ,\ell_o) \, . \label{v_k}
\ee
The radius $r_o$ and {the strength of the} kick, $\dot\theta_k$ are two initial parameters, from which the evolution of the particle follows.
Later, we will provide a complete analysis to determine the type of the final states for
 various choices of $r_o$ and $\dot\theta_k$.
 Since the kick is along the $\theta$ direction,
the particle's  azimuthal angular momentum remains the same value as $\ell_o$ in~(\ref{l_o}).
The  energy $\varepsilon$ on the other hand will be modified  after a kick and  the normalization condition for the  $u_{i \, \nu}$ gives
\be \label{energyk}
\varepsilon = \frac{\ell_o a(2Mr_o-Q^2)+r_o\Delta_o^{1/2}\sqrt{(r_o^4+a^2(r_o^2+ 2 M r_o-Q^2))(1+r_o^2 \dot{\theta_k^2})+r_o^2 \ell_o^2}\;}{r_o^4+a^2(r_o^2+ 2 M r_o-Q^2)} \, ,
\ee
where, $\Delta_o\equiv \Delta(r_{o})$. The {form} of $\varepsilon$ {is  chosen so that the four-velocity is future-directed.
In addition, {the Carter constant reduces to}
\be
\kappa =r_o^4 \dot\theta_k^2
\ee in~(\ref{thetadot}).

The effective potential defined in~(\ref{v_eff}) depends {on both the azimuthal angular momentum $\ell$ and the} energy $\varepsilon$.
So, the analysis of the particle's dynamics proceeds by rewriting  ~(\ref{rdotsquare}) as
\be
\Sigma^2 \dot{r}^2 = (r^4+a^2(r^2+ 2 M r-Q^2))(\varepsilon-V_+)(\varepsilon-V_-) \, \label{rdotdotvpvm}
\ee
where
\be \label{V_pm_k}
V_\pm (r)= \frac{\ell a(2Mr-Q^2) \pm \Delta^{1/2} \sqrt{a^2(\kappa +r^2)(r^2+ 2 M r-Q^2)+r^4(\kappa+r^2+\ell^2)}\;}{r^4+a^2(r^2+ 2 M r-Q^2)} \, .
\ee
The acceleration $\ddot{r}(r)$ can be obtained by taking the proper time derivative of~(\ref{rdotdotvpvm}), and {we find}
\be \label{V+}
\ddot{r}(r) = -\frac{r^4+a^2(r^2+ 2 M r-Q^2)}{2r^4}(\varepsilon-V_-(r))V_+'(r) \, .
\ee
Eq.~(\ref{V+}) shows that {when} $\varepsilon > V_-(r_o)$,  $\ddot{r}(r_o) \propto -V_+'(r_o)$,  where $r_o$ is the initial radius of the orbital motion.  Thus, $V_+ (r)$ is more relevant for the dynamics of the particle. Apparently, whether the particle potentially moves toward the spatial infinity or is captured by the black hole crucially depends on the sign of  $\ddot{r} (r_o) $. The curve {of the critical $\dot{\theta}$ that separates the two cases in the parameter space} is determined by
\be
V_+'(r_{o},\dot{\theta}_c) = 0  \, .
\ee
 The behavior is illustrated in Fig.~\ref{regions} for a critical kick  of $\dot\theta_c$. The region to the left of the critical curve corresponds to ${\ddot r} <0$ while the other side  corresponds to ${\ddot r} >0$.
The critical kick of $|\dot{\theta}_c|$ starts from zero  at $ r_o=r_{I}$ and approaches to infinity as $r_o$ approaches $r_F$.
For a given $a$ and $Q$, {$r_{I}$ and $r_{F}$ satisfy}
\begin{eqnarray} \label{r_IF}
&& a \, \sqrt{M r_I- Q^2 } \big[ r_I^2 ( 2 Q^2 + r_I ( 2 r_I - 3M))+ a^2 ( Q^2 + r_I ( 2 r_I -M ))\big] \nonumber\\
 && \quad\quad \quad\quad\quad \quad + r^4 ( a^2 + 2 Q^2 + r_I ( r_I - 3M)) + a^4 ( Mr - Q^2)=0 \, ,\\
&& a^2(r_{F}+M)+r_{F}(2Q^2+r_{F}(r_{F}-3M))=0 \, .
\end{eqnarray}
Thus, the values of $r_I$ and $r_F$
decrease as the value of $Q$ (and/or $a$) increases as shown in Fig.~\ref{dresccap1},
and they coincide when $a=0$  for any possible $Q$, also seen in the figure. These are the generalization of the findings in~\cite{ZAH1} to the Kerr-Newman black hole.
\begin{figure}
  \centering
  \includegraphics[scale=0.4]{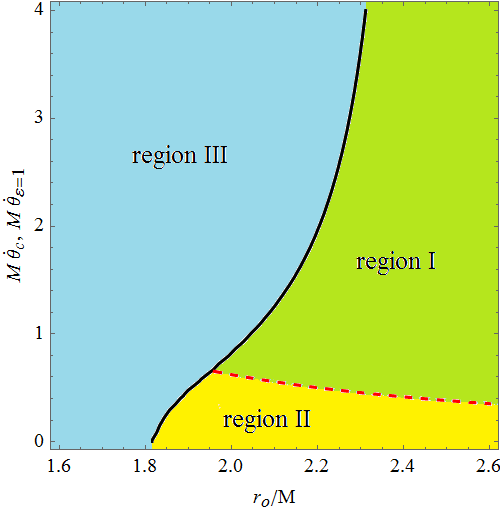}
\caption{The solid curve of $\theta_{c}$ that corresponds to $V_+'(r_{o},\dot{\theta}_c) = 0$ is plotted with respect to $r_o$ for $a=0.96M$ and $Q=0.2M$. $|\dot{\theta}_c|$ starts from zero  at $ r_o=r_{I}=1.82 M$ and approaches to infinity as $r_o$ approaches $r_F=2.43 M$. The red dashed curve describes $ |\dot{\theta}_{\varepsilon = 1}|$ when the energy of a particle is $\varepsilon=1$. } \label{regions}
\end{figure}
\begin{figure}
  \centering
  \includegraphics[scale=0.45]{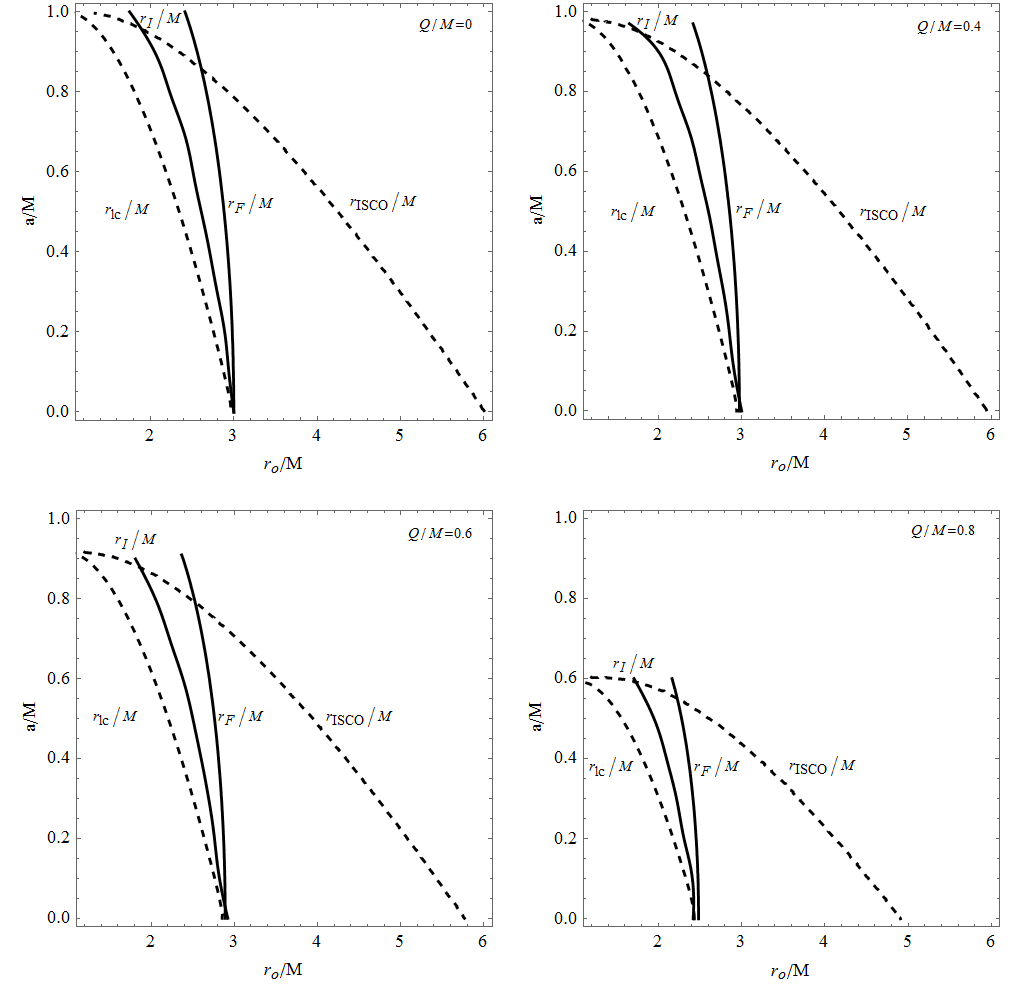}
\caption{The dependence of $r_{I}$ and $r_{F}$ on $a$ and $Q$ is shown. Also, $r_{lc}$ and $r_{\rm ISCO}$ are displayed  as the function of  $a$ and $Q$ for comparison. When $ r_o > r_{\rm ISCO}$, the stable initial circular motion exists, and when $r_{lc} < r_o < _{\rm ISCO}$, the unstable circular motion is possible.} \label{dresccap1}
\end{figure}
In Fig.~\ref{dresccap1}, the $r_{\rm ISCO}$ is also shown for various choices of $a$ and $Q$.
  With  an initial $r_o > r_{\rm ISCO}$ and a given value of $\dot\theta_k$ of a kick, one  can determine the possible sign of $\ddot r$ from Fig.~\ref{regions}.
Nevertheless, for $r_{lc} < r_o < r_{\rm ISCO}$ where $R''(r_o)<0$, the particle can  be initially in unstable circular motion. The radius of the last circular orbit $r_{lc}$, at which  both $\varepsilon_o$ in (\ref{e_o}) and $\ell_o$ in (\ref{l_o}) become infinity, is  found by
\be \label{r_lc}
2 a (M r_{lc}-Q^2)^{1/2}+2Q^2+r_{lc}(r_{lc}-3M) = 0 \,
\ee
and is shown in Fig.~\ref{regions} for  comparison.
{When} $\ddot r >0$, the particle {may} escape to spatial infinity {or be bounded, depending on whether its energy $\varepsilon$ is greater than one or not}. On the other hand, for $\ddot r <0$, the particle is mainly captured by the black hole.
Thus, we will consider these cases separately below.
As for negative value of  $a$ for the counter-rotation motion for the particle around a Kerr-Newman black hole, since $V_{\pm} (-a) =- V_{\mp} (a)$ in~(\ref{V_pm_k}), the same
strategy for $a>0$ case  can be straightforwardly applied~\cite{Stuchlik:2008fy}.

In the regime of the parameters ($\dot\theta_k, r_o$) with  $\ddot r >0$, the important criterion to determine whether the particle will escape to spatial infinity or stay in the bounded motion is set by the condition $\varepsilon=1$ in~(\ref{energyk}).
The {corresponding value} of the kick $\dot\theta_{\epsilon=1}$ is obtained by
\be \label{theta_c}
|\dot{\theta}_{\varepsilon = 1}| = \bigg[\frac{(Q^2-2Mr_o)(a^2+r_o^2-2a\ell_o)+(\Delta_o-a^2)\ell_o^2}{r^4\Delta}\bigg]^{1/2} \, ,
\ee
and shown in Fig.~\ref{regions}.
{Inside regime (I) in Fig.~\ref{regions}, since the impulse force is adequately strong or the orbital radius is sufficiently large}, the particle then has large enough energy to escape from the black hole to spatial infinity.  This behavior can also be understood from the potential energy $V_{+}$ in Fig.~4a (regime (I)). However, in the regime (II) $\varepsilon<1$, the corresponding energy potential shows that the particle still get trapped around the local minimum of the effective potential (Fig.~4b), and remains the bounded motion with slightly larger radius as compared with the initial radius~\cite{Stuchlik:2008fy}.
On the contrary, for ($\dot\theta_k, r_o$) corresponding to} an initial circular motion  in the regime (III),  the particle with $\ddot r<0$ will move toward the black hole, and finally gets captured. This evolution is also seen by the potential energy (Fig.~4c) in the regime (III).
 All these behaviors will be finally justified by numerically solving the equations of motion of the particle.
\begin{figure}
  \centering
  \includegraphics[scale=0.3]{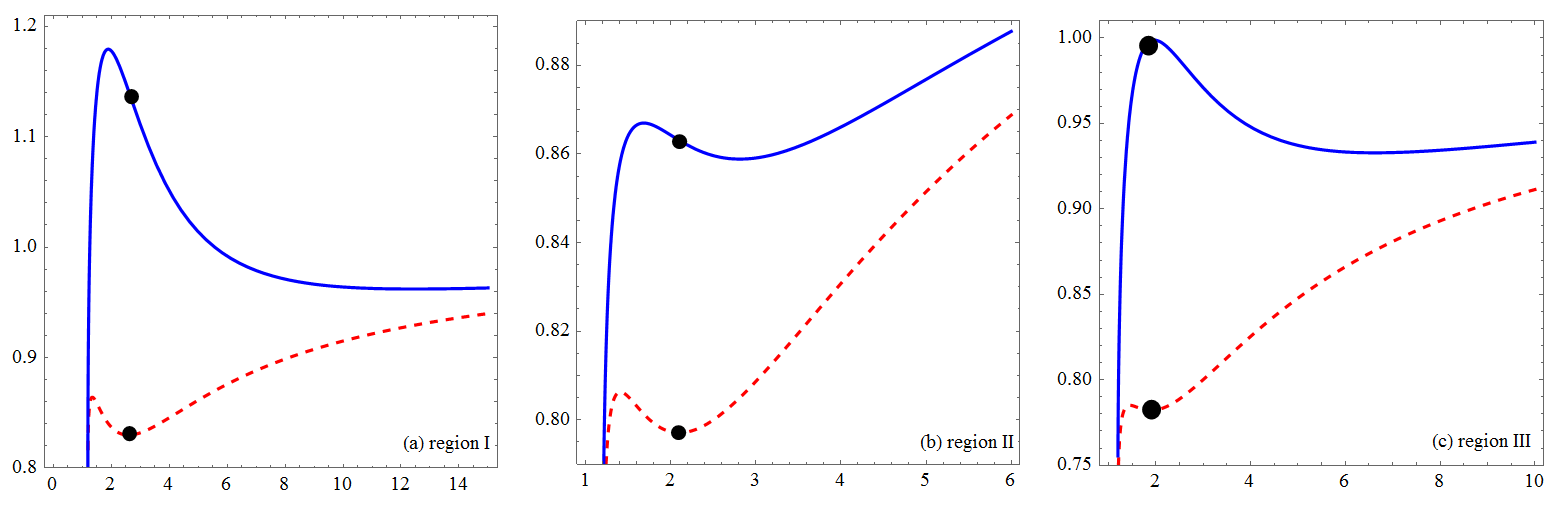}
\caption{
Illustrations of $V_+(r)$ for a particle in the black hole geometry of $ a = 0.96M $ and $Q = 0.2M$
  before (dashed red line) and after (solid blue line) a kick with the initial parameters in the regime
  (I) (a): $r_o=2.6M, \dot\theta_k=0.5/M $ ; $\ddot{r}>0, \varepsilon >1$,  (II) (b): $ r_o=2.1M, \dot\theta_k=0.3/M$; $\ddot r> 0, \varepsilon <1$, and (III) (c): $r_o=1.9M, \dot\theta_k=0.7/M$ ; $\ddot r<0$. The black dots on the dashed red line and the solid blue line indicate the respective position of the test particle before and right after the kicks. }     \label{region123}
\end{figure}

\section{Trajectories of the particle from solving its dynamical equations in the Kerr-Newman black hole}\label{sec4}
The dynamics of a neutral particle of mass $m$ in curved spacetime is governed by the {geodesic} equation
\be
m u^{\nu} \nabla_{\nu} u^{\mu} = 0 \, . \label{eqom}
\ee
In the Kerr-Newman black hole geometry, the equations of motion of the particle for the radial and polar components are
\begin{eqnarray}
\ddot{r} &=& (\Delta r \dot{\theta}^2+ a^2  \sin2\theta\,  \dot{r}\, \dot{\theta})/\Sigma+ \big[
{X Y}(aZ\sin^2\theta-\frac{Y}{4})+ {G Z^2}\sin^4\theta \big]/\Delta \Sigma^5 \, , \label{rdotdot} \\
\ddot{\theta} &=& (a^2 \sin^2\theta (\Delta \dot\theta -{\dot r}^2)- 4 \Delta r {\dot r} \dot\theta)/2 \Sigma \Delta
+\big[ {J Y}+ {H Z^2}\cos\theta \sin^5\theta + \frac{a^2 \sin2\theta  Y^2}{8}(Q^2+2Mr)\big]/\Delta \Sigma^5 \, \nonumber\\
\label{thetadotdot}
\end{eqnarray}
{accompanied} with two conserved quantities, $\varepsilon$ (\ref{energyk})  and $\ell_o$~(\ref{l_o}), to be determined by the initial radius of the orbit $r_o$
and a kick $\theta_k$. {Here} $X, Y, Z, P, G, J, H$ are {shorthand} notations for
\begin{eqnarray}
X &=& r(Q^2+Mr)-a^2 M\cos^2\theta \, , \nonumber\\
Y &=& 2a\ell(Q^2-2Mr)+a^4\varepsilon+2r^4\varepsilon+a^2\varepsilon(2Mr+3r^2-Q^2)+a^2\Delta\varepsilon\cos2\theta \, , \nonumber\\
Z &=& a(a\ell+\varepsilon(Q^2-2Mr))-\ell\Delta/\sin^2 \theta \, , \nonumber\\
P &=& r(a^2+Q^2-Mr)+a^2(M-r)\cos^2\theta \, , \nonumber\\
G &=&  a^4(M-r)\cos^2\theta +r(a^2(a^2+Q^2-Mr)+2a^2) (a^2+r^2) \cos^2 \theta/\sin^2 \theta+(r^4-a^4)/\sin^2\theta) \, , \nonumber\\
J &=&{16 a(a^2+r^2)
  (Q^2-2 M r) }\cos\theta/\sin\theta \nonumber\\
  &\times& (2
   \ell(\Delta
   -a^2)+a^2 \ell-a
   \varepsilon (Q^2-2 M r)+a \cos (2
   \theta )(a
   \ell+\varepsilon(Q^2-2
   M r))) \, ,  \nonumber\\
 H &=& -a^4\Delta+r^2(a^2+r^2)^2/\sin^4\theta+a^2  (a^4-r^2
  (-4 M r+2 Q^2+r^2))/\sin^2\theta \nonumber \\
  && +(a^2 (a^2+r^2)^2/\sin^2\theta-2 a^4 \Delta )\cos^2\theta/\sin^2\theta \, .
\end{eqnarray}
We then solve~(\ref{rdotdot}) and ~(\ref{thetadotdot}) numerically for a neutral particle around the black hole ($a/M = 0.9$, $Q/M = 0.2$) directly in the proper time.  For the parameters in the regime I, with  $\ddot r>0$ and the energy $\varepsilon >1$, the effective potential Fig.~4a shows that  the particle of sufficiently large energy can escape to spatial infinity with a trajectory in Fig.~5a. As for the parameters in the regime II, since the kick does not give the particle sufficient energy ($\varepsilon<1$, $\ddot r >0$), the particle remains the bounded motion as seen from the effective potential in Fig.4b with a trajectory in Fig. 5b. However, in the regime of the parameters III, the corresponding effective potential is shown in Fig. 4c with acceleration toward the black hole. Thus, the particle finally falls into the black hole in a trajectory in Fig.~5c.
\begin{figure}
  \centering
  \includegraphics[scale=0.3]{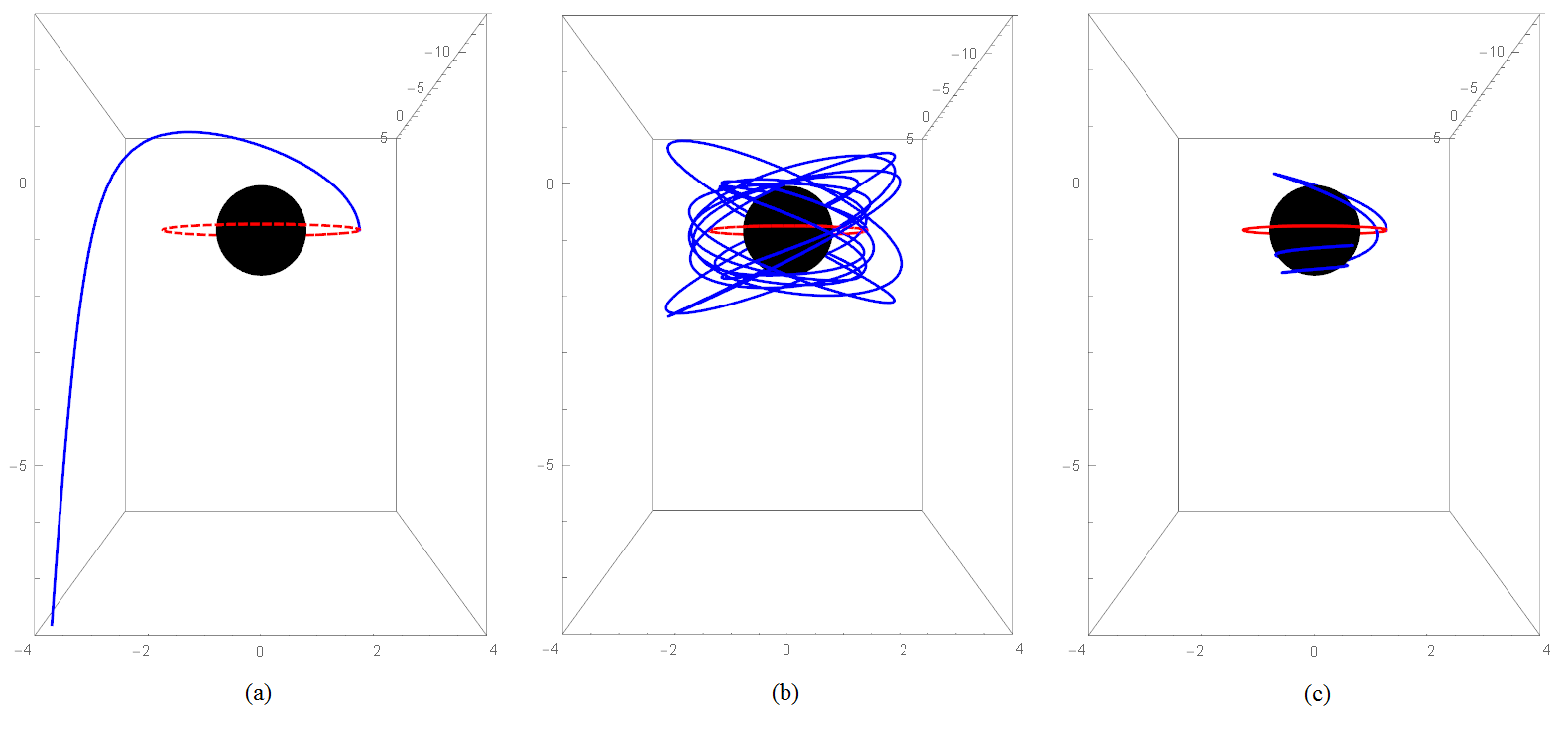}

  \caption{Demonstrations of trajectories for a particle  in the black hole geometry of $ a = 0.96M $ and $Q = 0.2M$ with the same parameters as in~(\ref{region123})  in the regime
  (I) (a): $r_o=2.6M, \dot\theta_k=0.5/M $ ; $\ddot{r}>0, \varepsilon >1$,  (II)  (b): $ r_o=2.1M, \dot\theta_k=0.3/M$; $\ddot r> 0, \varepsilon <1$, and (III) (c): $r_o=1.9M, \dot\theta_k=0.7/M$ ; $\ddot r<0$.  The black region represents interior of the horizon, and the circle (red) means the initial orbit of the particle.} \label{Traj123}
\end{figure}

\begin{figure}
  \centering
  \includegraphics[scale=0.4]{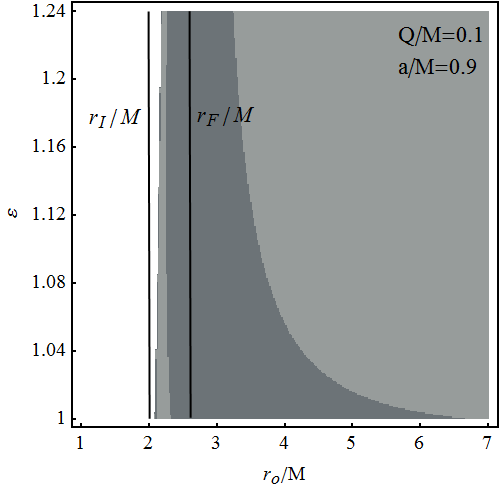}

  \caption{The basin of attraction plot for a neutral particle
   when $Q/M=0.1$ and $a/M=0.9$. $r_I$ and $r_F$ are shown as well. See the text for detailed descriptions on the shaded (blank) zones in parameter space ($r_o, \varepsilon$). } \label{BOAEX}
\end{figure}

\begin{figure}
  \centering
  \includegraphics[scale=0.3]{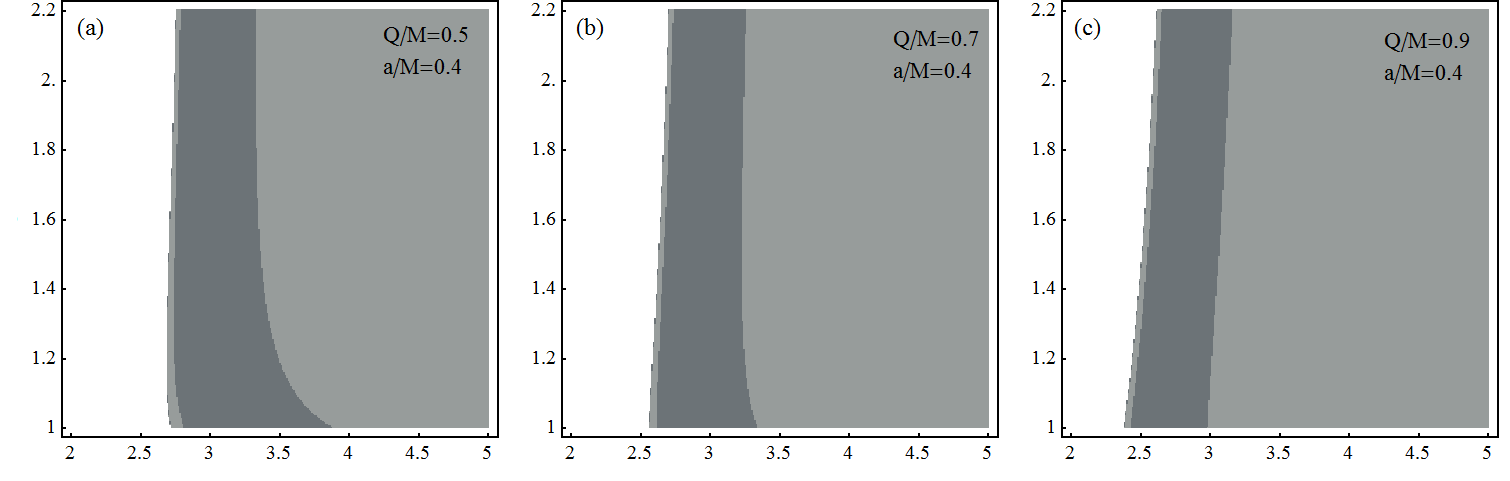}
  \includegraphics[scale=0.3]{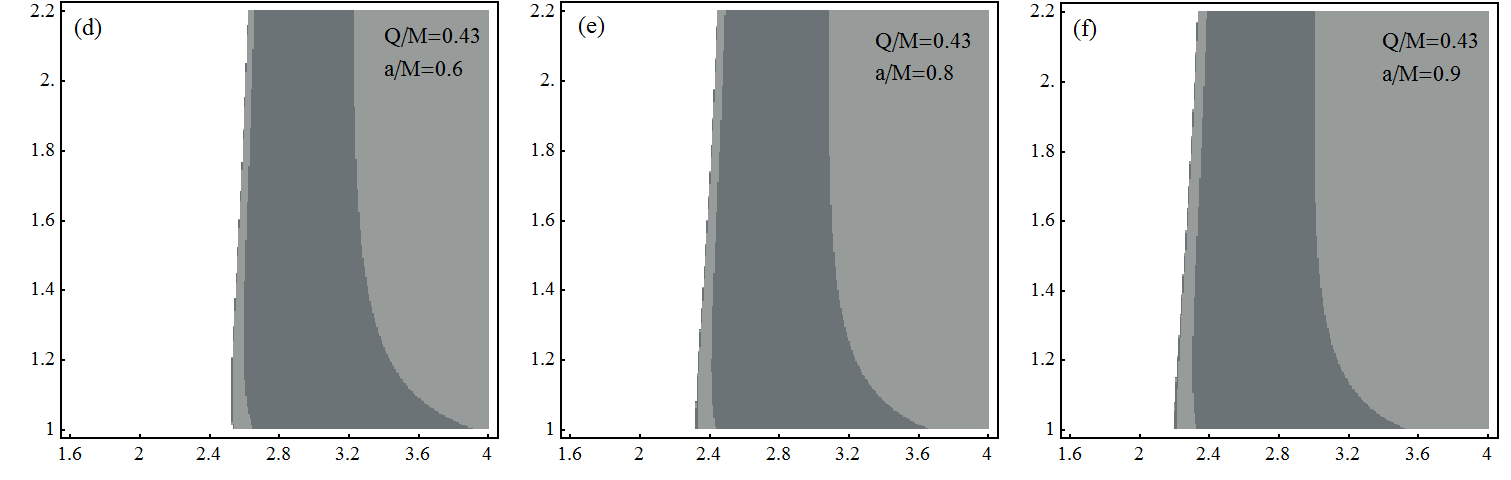}
\includegraphics[scale=0.3]{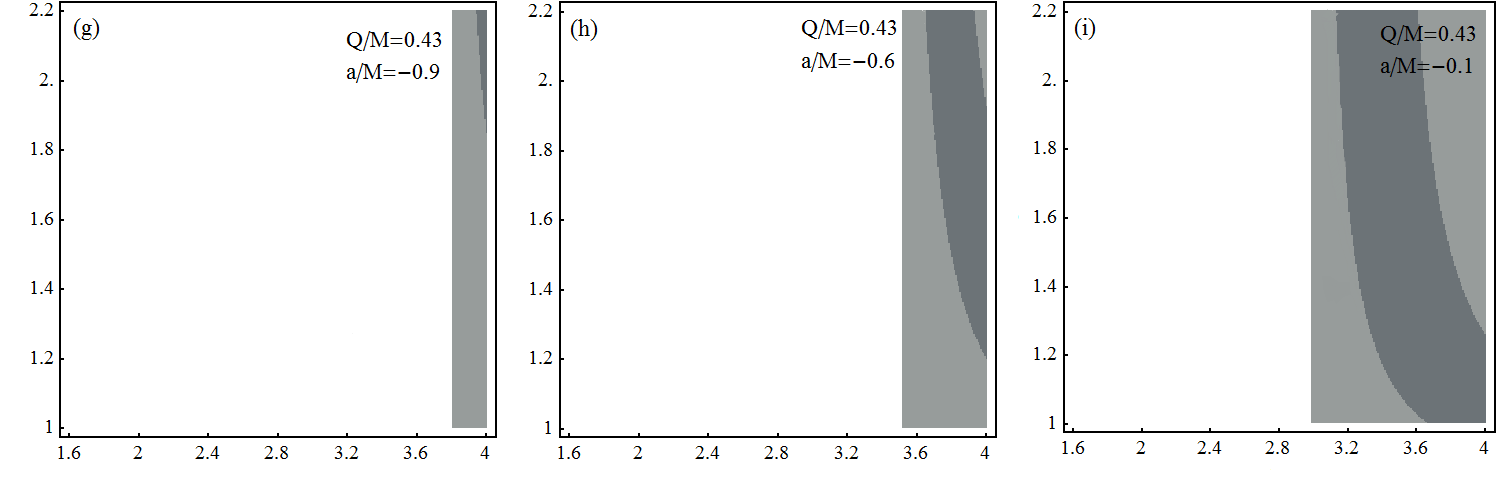}
  \caption{The basin of attraction plot for a neutral particle in parameter space ($r_o,\varepsilon$) as in Fig.~(\ref{BOAEX}), when  $ (a): Q/M=0.5, (b): Q/M=0.7, (c) Q/M=0.9$ for a fixed $a/M=0.4$; and $(d):a/M=0.6,(e): a/M=0.8, (f): a/M=0.9$, $(g):a/M=-0.9,(h): a/M=-0.6, (i): a/M=-0.1$ for a fixed $Q/M=0.43$.
   } \label{BOAQ043a060809}
\end{figure}

An extensive numerical study is employed with the method of basin of attraction. The idea of basin of attraction is
to identify the sets of initial values, from which the dynamical system tends to asymptotically evolve into an attractor.
As to astrophysical observations of interest, we will restrict the initial conditions with the energy $\varepsilon >1$, so that
the {particle} may either get captured by the black {hole}, or escape to spatial infinity.

The plot of basin of attraction for the black hole $Q/M=0.1$ and $a/M=0.9$ is shown in Fig.~\ref{BOAEX}  with the initial radius of the orbit motion and the energy after a kick in the regions  $r_{lc}=1.54 M < r_o <7 M$,  and $1<\varepsilon <1.24$. Notice that the radius of the inner most stable circular motion, the last circular motion (\ref{r_lc}), and the black hole horizon are
$r_{\rm ISCO}=2.29 M$, $r_{lc}=1.54 M$, and
$r_H=1.42 M $ for reference.
The maxima integration time is about $\tau= 10^4 M$.
 The particle is captured by the black hole  with initial conditions lying in the blank area, and escape to spatial infinity in the grey area. In particular, the direction of the escape is opposite to (same as)  the initial kick with the $(r_o,\varepsilon)$ in the light(dark) grey area. The radius of $r_I=2.03M $ and $r_F=2.55M $ in (\ref{r_IF}) are also plotted. For $\varepsilon >1$, the particle will fall into the black hole for $r_o < r_I$, and escape to spatial infinity for $r_o > r_F$ as seen in the graph.  The result of~\cite{ZAH} can be reproduced in the $Q=0$ case.
{ In Fig.~7a-7c, the blank area of $(r_0, \varepsilon)$  to the capture  decreases  as the value of $Q$ increases because the increase of $Q$ will lead to larger repulsive effects to the dynamics of the particle as also seen in~(\ref{v_eff}).
 Along the same line of thoughts, the capture region in Fig.~7d-7f increases as $a$ ($ a>0$, for co-rotation) decreases or the absolute value of $a$ ($ a<0$, for counter-rotation) increases, also seen in Fig.~7g-7i.  Moreover, numerical calculation indicates that the basin boundaries are regular lines, as  predicted by the analytical results in~(\ref{theta_c}). They should be the regular motion with no chaotic behavior. Thus, the chaoticness of the motion might be seen when the particle under consideration carries a charge
 with the inclusion of the magnetic fields~\cite{ZAH},  which deserves further study.

\section{summary and outlook}\label{sec5}

{In summary, the dynamics of a neutral particle around a Kerr-Newman black hole is studied with emphasis on how the electromagnetic
fields intrinsically generated from the spinning charged black hole influences the particle's dynamics.
We first {examine} the innermost stable circular orbits on the equatorial plane. It is found that the presence of the charge of the black {hole} gives the  effective potential of the particle with stronger repulsive effects as compared with the Kerr back {hole}. As a result, the radius of ISCO decreases as the value of $Q$ increases for a fixed angular momentum of the black hole $a$.
We consider {an impulse that kicks}  the particle which initially moves around an orbit, out of the equatorial plane.
After a kick, the particle may remain the bounded motion, escape to spatial infinity, or fall into the black hole.
We then provide the analytical and numerical studies {to identify the regions in the parameter space which lead} to the above respective final states.
 In particular, {when the particle is captured by the black hole, the corresponding region in the parameter space} decreases as $Q$ increases.
Additionally, the boundary of the parameters $(\dot\theta_k, r_o$) between the capture and {runaway} regions {appears as} a regular {curve}.
The chaotic motion thus may be expectedly seen for the motion of a charged particle in the external magnetic field~\cite{ZAH}.
Our work certainly provides the first step toward understanding the dynamics of the particle moving in the {spacetime that has} a spinning charged black hole.  {In the next step} we will consider the motion of the charged particle to explore its chaotic behavior in the Kerr-Newman black hole with and/or without an external magnetic field. It would be of great interest to compare with  the chaotic phenomena of a charged particle in the magnetized Kerr black hole.

\begin{acknowledgments}
 This work was supported in part by the
Ministry of Science and Technology, Taiwan.
\end{acknowledgments}


\begin{thebibliography}{99}
\bibitem{wheeler}
C. W. Misner, K. S. Thorne, and J. A.  Wheeler, {\it  Gravitation} (  W. H. Freeman and Company,  San Francisco, 1973).

\bibitem{hartle}  J.  B. Hartle, {\it Gravity: An Introduction to Einstein's General Relativity} (Addison-Wesley, 2003).
%
\bibitem{Aliev:2005bi}
  A.~N.~Aliev and A.~E.~Gumrukcuoglu,
  Phys.\ Rev.\ D {\bf 71}, 104027 (2005).
%
\bibitem{Stuchlik:2008fy}
  Z.~Stuchlik and A.~Kotrlova,
  Gen.\ Rel.\ Grav.\  {\bf 41}, 1305 (2009).
%
\bibitem{Blaschke:2016uyo}
  M.~Blaschke and Z.~Stuchlík,
  Phys.\ Rev.\ D {\bf 94}, no. 8, 086006 (2016).
%
\bibitem{Pugliese:2011xn}
  D.~Pugliese, H.~Quevedo and R.~Ruffini,
  Phys.\ Rev.\ D {\bf 84}, 044030 (2011).
%
\bibitem{RUF2} D. Pugliese, H. Quevado, and R. Ruffini, Phys. Rev. D {\bf 83},024021 (2011).
%
\bibitem{RUF1}  D. Pugliese, H. Quevado, and R. Ruffini, Phys. Rev. D {\bf 83},104052 (2011).


\bibitem{Nara}
R. Narayan and J. McClintock, Mon. Not. R. Astron. Soc.
{\bf 419}, L69 (2012).

\bibitem{Tch}
A. Tchekhovskoy, R. Narayan and J. C. McKinney, Mon. Not. Roy. Astron. Soc. {\bf 418}, L79 (2011).

\bibitem{FRO1}
V. P. Frolov and A. A. Shoom,  Phys. Rev. D {\bf 82}, 084034 (2010).

\bibitem{FRO2}
V. P. Frolov, Phys. Rev. D
{\bf 85}, 024020 (2012).

\bibitem{ZAH}M. Al Zahrani, Valeri P. Frolov and Andrey A. Shoom, Phys. Rev. D. {\bf 87}.084043 (2013).

\bibitem{ZAH1}A. M. Al Zahrani. Phys. Rev. D {\bf 90}, 044012 (2014).

\bibitem{RYOS}
R. Shiose, M. Kimura, and T. Chiba, Phys. Rev. D {\bf 90}, 124016 (2014).
\bibitem{NAK}
Y. Nakamura and T. Ishizuka, Astrophys. Space Sci. {\bf 210}, 105 (1993).

\bibitem{TAK}
M. Takahashi and H. Koyama, Astrophys. J. {\bf 693},
 472 (2009).

\bibitem{HUS} S. Hussain, I. J. M. Hussain,  Eur. Phys. J. C {\bf 74}, 3210 (2014).

%
\bibitem{Damo}
T. Damour, R. Hanni, R. Ruffini, and J.Wilson, Phys. Rev.
D {\bf 17}, 1518 (1978).
%
\bibitem{Pugliese:2013zma}
  D.~Pugliese, H.~Quevedo and R.~Ruffini,
  Phys.\ Rev.\ D {\bf 88}, 024042 (2013).
%

\bibitem{RUF4}
M. Johnston and R. Ruffini, Phys. Rev. D,
{\bf 10},2324  (1974).

\bibitem{Young}
P. J. Young, Phys. Rev. D 14, {\bf 3281} (1976).

\bibitem{Sharp}
N. A. Sharp, Gen. Relativ. Gravit. 10, {\bf 659} (1979).

\bibitem{Bicak1}
J. Bicak, Z. Stuchlik, and V. Balek, Bull. Astron. Inst.
Czech. {\bf 40}, 65 (1989).

\bibitem{Bicak2}
V. Balek, J. Bicak, and Z. Stuchlik, Bull. Astron. Inst.
Czech. {\bf 40}, 133 (1989).

\bibitem{Bicak3}
Z. Stuchlik, J. Bicak, and V. Balek, Gen. Relativ. Gravit.
{\bf 31}, 53 (1999).

\bibitem{Kovar}
J. Kovar, Z. Stuchlik, and V. Karas, Classical Quantum
Gravity 25, 095011 (2008).

\bibitem{Hagi}
Y. Hagihara, Jpn. J. Astron. Geophys. {\bf 8}, 67 (1931).

\bibitem{Gru}
S. Grunau and V. Kagramanova, Phys. Rev. D {\bf 83}, 044009
(2011).

\bibitem{Hack}
E. Hackmann and H.  Xu, Phys. Rev.  D {\bf 87}, 124030 (2013).
%
\bibitem{Mino} Y. Mino, Phys. Rev. D {\bf 67}, 084027 (2003).

\end{thebibliography}
\end{document}